\documentclass[twocolumn,amsmath,amssymb,prl]{revtex4}
\usepackage{graphicx}% Include figure files
\usepackage{dcolumn}% Align table columns on decimal point
\usepackage{bm}% bold math
\begin{document}

%\baselineskip=2\normalbaselineskip %this is good for the corrections

% Use the \preprint command to place your local institutional report
% number in the upper righthand corner of the title page in preprint mode.
% Multiple \preprint commands are allowed.
% Use the 'preprintnumbers' class option to override journal defaults
% to display numbers if necessary
%\preprint{Marques et al.}

%Title of paper
\title{A possible mechanism for cold denaturation of proteins \\
 at high pressure}

% repeat the \author .. \affiliation  etc. as needed
% \email, \thanks, \homepage, \altaffiliation all apply to the current
% author. Explanatory text should go in the []'s, actual e-mail
% address or url should go in the {}'s for \email and \homepage.
% Please use the appropriate macro foreach each type of information

% \affiliation command applies to all authors since the last
% \affiliation command. The \affiliation command should follow the
% other information
% \affiliation can be followed by \email, \homepage, \thanks as well.
\author{Manuel I.  Marqu\'es$^1$}
\email{manuel@argento.bu.edu}
\author{Jose M. Borreguero$^1$}
\author{H. Eugene Stanley$^1$}
\author{Nikolay V. Dokholyan$^2$}

\affiliation{%
$^1$~Center for Polymer Studies and Department of Physics,
Boston University, Boston, MA 02215, USA\\
$^2$~Department of Biochemistry and Biophysics,
University of North Carolina at Chapel Hill, School of Medicine,
Chapel Hill, NC 27599, USA
}%

%Collaboration name if desired (requires use of superscriptaddress
%option in \documentclass). \noaffiliation is required (may also be
%used with the \author command).
%\collaboration can be followed by \email, \homepage, \thanks as well.
%\collaboration{}
%\noaffiliation

%\date{\today}

\begin{abstract}
  We study cold denaturation of proteins at high pressures. Using
  multicanonical Monte Carlo simulations of a model protein in a water bath,
  we investigate the effect of water density fluctuations on protein
  stability. We find that above the pressure where water freezes to the dense
  ice phase ($\approx2$~kbar), the mechanism for cold denaturation with
  decreasing temperature is the loss of local low-density water structure. We
  find our results in agreement with data of bovine pancreatic \mbox{ribonuclease A}.
\end{abstract}

% insert suggested PACS numbers in braces on next line
%\pacs{}
% insert suggested keywords - APS authors don't need to do this
%\keywords{}

%\maketitle must follow title, authors, abstract, \pacs, and \keywords
\maketitle

% If in two-column mode, this environment will change to single-column
% format so that long equations can be displayed. Use
% sparingly.
%\begin{widetext}
% put long equation here
%\end{widetext}

Some proteins become thermodynamically unstable at low temperatures, a
phenomenon called cold denaturation \cite{Pace,Privalov,Jonas}. This
phenomenon has been mainly observed at high pressures, ranging from
approximately 200 MPa up to 700 MPa \cite{Kunugi}.  An explanation of the
$P-T$ phase diagram of a protein with cold denaturation has been proposed
\cite{Hawley}, but a microscopic understanding of the mechanisms leading to
cold denaturation has yet to be developed, due in part to the complexity of
protein-solvent interactions.

Existing theories of folding and unfolding of diluted proteins consider
hydrophobicity as the driving force of protein stability
\cite{Kauzmann,Dill,DillII,Hummer98,Chandler}.  In the case of apolar
macromolecules, hydrophobicity has been identified with the assembly and
segregation of the macromolecule to minimize the disruption of hydrogen bonds
among water molecules \cite{Kauzmann,Chandler,Stillinger}.  Water tends to be
removed from the surface of apolar molecules, forming a cage composed of
highly organized water molecules around the molecule, where the disruption of
hydrogen bonds is minimized \cite{Frank}. The simplest hydrophobic model
features an effective attraction between hydrophobic
molecules~\cite{Vanderzande}, but does not reproduce cold denaturation. In
order to obtain cold denaturation with this model, new studies \cite{Dill89,
  Shimizu02} had to insert a temperature-dependent attraction derived from
experimental observations at ambient pressure \cite{Nozaki71}. An explicit
account of water around the hydrophobic molecules has also been considered in
order to understand the cold denaturation process with
temperature-independent interactions. Theoretical attempts modeled the
effective water-protein interactions with the free energy cost of excluding
the solvent around the nonpolar molecule \cite{Stillinger,Pratt}.  Numerical
simulations based on these attempts have been applied to study the pressure
denaturation found in proteins \cite{Hummer98}.

Not until recently has cold denaturation been studied at the microscopic
level.  Simplified models \cite{Rios}, based on a bimodal description of the
energy of water in the shell around the hydrophobic molecule \cite{Muller},
predicted cold denaturation. Similar results were obtained using a lattice
model of a random hydrophobic-hydrophilic heteropolymer interacting with the
solvent \cite{Trovato}. Several models mimicking the interaction between
water molecules and non-polar monomers have also been applied to the study of
cold denaturation.  \cite{Bakk}. One possible reason for the inability of the
previous models to capture both the molecular details of cold denaturation
and the effect of pressure is the neglect of (i) correlations among water
molecules near the freezing point, and (ii) the density anomaly due to the
tetrahedral structure of the hydrogen bonded network.  Here, we implement a
two-dimensional lattice model of water that captures the above mentioned
water properties \cite{FranzeseII}. In the model, the possible orientations
of water molecules are set by the allowed values of a $q$-state Potts
variable $\sigma_{i}$.  Only when two neighbor molecules $\langle i,j\rangle$
are in the correct orientation ($\sigma_{i}=\sigma_{j}$) does a hydrogen bond
(HB), that increases the volume of the system by $\Delta V$, form. This
interaction mimics the increment of volume due to the incipient formation of
a tetrahedral structure. If the two neighbor molecules $\langle i,j\rangle$
are not in the correct orientation, the interaction of the particles does not
imply any increment in volume. The Hamiltonian for our model of water-water
interaction may be written as \cite{FranzeseII}

\begin{equation}
{\cal H}_{HB}=
    -J\sum_{\langle i,j\rangle}{\delta_{\sigma_{i},\sigma_{j}}} ~,
\label{eq1}
\end{equation}
were $J>0$ is the scale of the interaction between water molecules upon
tetrahedral network formation.  The total volume of the system is given by
$V=V_{0}+N_{HB}\Delta V$, where $N_{HB}= \sum_{\langle i,j\rangle}{\delta_
  {\sigma_{i},\sigma_{j}}}$ is the total number of hydrogen bonds with
$\Delta V>0$ in the system. The sum $\sum_{\langle i,j\rangle}$ extends
  only to nearest neighbors, implying that two water molecules cannot form a
  hydrogen bond with $\Delta V >0$ if they are separated by one residue of
  the protein. The enthalpy of the system (${\cal H}_{HB}+PV$) is given by

\begin{equation}
{\cal H}_{HB}+PV=
    -(J-P\Delta V)\sum_{\langle i,j\rangle}{\delta_{\sigma_{i},\sigma_{j}}} ~,
\label{eq2}
\end{equation}
where $P$ is the pressure applied to the system.

Our model solvent features a limiting pressure $P_c = J/\Delta V$. Above
$P_c$, we find that $N_{HB}$ decreases as we decrease the temperature, and
the water model undergoes a transition to a state where all hydrogen bonds
with $\Delta V>0$ are broken. Below $P_c$, we find that $N_{HB}$ increases as
we decrease the temperature, and the water model undergoes a sharp
transition, at $T=T_c=(J-P\Delta V)/(\ln(1+\sqrt{q}))$ \cite{FranzeseII}, to
a state where all hydrogen bonds with $\Delta V>0$ are formed. Thus, our
water model reproduces the freezing of water to low and high density ice,
since for $P<P_c$ ($P_c \approx 200$MPa in real water) water freezes to the
low density ice Ih, and for $P>P_c$, water freezes to the high density ice II
\cite{Petrenko}. A relation between these two phases of ice and protein
folding has already been suggested from a thermodynamic point of
view~\cite{Wilse}.

We model the protein as a self-avoiding random walk embedded in a water bath.
For simplicity, we consider a non-polar homopolymer that interacts with water
via the partial ordering of water molecules, forming hydrogen bonded
structures around the protein. We mimic the interaction using the
Hamiltonian,

\begin{equation}
{\cal H}_{p}=
    J_{r}n_{HB}(n_{max}-\sum_{\langle i,j\rangle}{n_{i}n_{j}}) ~,
\label{eq3}
\end{equation}
where the parameter $J_{r}>0$ is the strength of the repulsive interaction
and $n_{HB} \equiv N_{HB}/N_{water}$ is the number density of hydrogen bonds
with $\Delta V>0$, $N_{water}$ is the number of water molecules. The
water--protein repulsion increases as the water molecules tend to form the
tetrahedral, low--density, hydrogen bonded network, where an unfolded apolar
macromolecule is unlikely to be embedded.  $n_{max}$ is the maximum number of
residue--residue contacts and $\sum_{\langle i,j\rangle}{n_{i}n_{j}}$ is the
number of residue--residue contacts, where $n_{i}=1$ if the lattice position
$i$ is occupied with a residue, and zero otherwise. Therefore,
$n_{max}-\sum_{\langle i,j\rangle}{n_{i}n_{j}}$ is a measure of protein
compactness, and equals zero when the protein is maximally compact. Thus,
Eq.~(\ref{eq3}) states that the hydrophobic repulsion driving the protein to
a compact state is equal to zero when the protein is maximally compact
$(\sum_{\langle i,j\rangle}{n_{i}n_{j}} \approx n_{max})$ or when the water
forms the high--density bond network ($n_{HB}\approx0$).

We hypothesize that the inability of water molecules to arrange in the low
density ice--like structures is the principal mechanism responsible for
protein cold denaturation.  At low pressures ($P<P_c$) and low temperatures,
water molecules form a low density hydrogen bonded network, so the protein is
forced to adopt a compact state. At high pressures ($P>P_c$), the water is
not able to form the low density network and forms a more dense state.  In
this case the effective repulsion between the residues and the solvent
decreases, and water molecules penetrate into the protein core, unfolding the
compact state. Our hypothesis is supported by the experimental observations
\cite{Kunugi} that cold denaturation exists mainly at high pressures (of the
order of kbars), where water only freezes in the dense ice II
phase~\cite{other}.

\begin{figure}
\includegraphics[width=7cm,height=7cm,angle=-90]{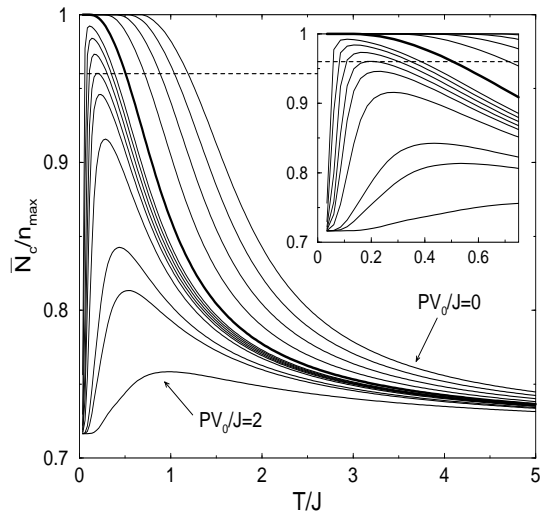}
\caption{
 Normalized number of residue-residue contacts vs. temperature for pressure
 values $PV_{0}/J=0,0.25,0.5,
 0.75,1,1.1,1.125,1.15,1.175,1.2,1.25,1.4,1.5,2$. A magnified region where
 the cold denaturation takes place is shown in the inset. The dotted line
 corresponds to $\overline N_{c}/n_{max}>0.96$. We represent the curve
 corresponding to $P=P_c=J/\Delta V$ by a bold line. The values of the
 parameters used are: $J=1, J_{r}=10, \Delta V=1$ and $q=10$.  }
\label{fig1}
\end{figure}

Next we demonstrate that our model of a protein embedded in a water network
with an enthalpy given by

\begin{equation}
{\cal W}={\cal H}_{p}+{\cal H}_{HB}+PV 
\label{eq4}
\end{equation}
gives rise to both cold and warm denaturation of the protein and agrees with
experimentally-observed protein denaturation at high pressures.  Since the
energy landscape of the protein interacting with the water network is
characterized by a multitude of local minima, we perform multicanonical Monte
Carlo simulations to avoid transient trapping of our water-protein system in
local energy minima at low temperatures. Specifically, we use the
multiple-range random walk algorithm~\cite{Wang} to calculate the density of
states. We adopt the algorithm in order to embed the self-avoiding protein
into the lattice, and to calculate the two-parameter density of states
$g(N_{HB},N_{c})$, where $N_{c}\equiv\sum_{\langle i,j\rangle}{n_{i}n_{j}}$
is the number of residue-residue contacts. From the density of states we
calculate the temperature and pressure dependence of the average number of
residue-residue contacts

\begin{equation}
{\overline N_{c}}=\sum_{N_{HB}}\sum_{N_{c}}N_{c}g(N_{HB},N_{c})\frac {e^{-{\cal W}(N_{HB},N_{c})/T}}{Z} ,
\label{eq5}
\end{equation}
where $Z$ is the partition function. We perform Monte Carlo simulations of a
system of $383$ water molecules and a protein consisting of $17$ non-polar
residues with periodic boundary conditions.  Fig.~\ref{fig1} shows the
dependence of $\overline N_{c}/n_{max}$ on temperature for different values
of the pressure both above and below $P_{c}$.  The calculated density of
states $g(N_{HB},N_{c})$ converges to the true value with an accuracy of the
order of $10^{-5}$. The value of $\overline N_{c}/n_{max}$ ranges from one
(maximally compact protein) to approximately $0.71$ (which is the average
number of residue--residue contacts found at high temperatures). Only when
$P>P_c$ do we observe the cold denaturation of the protein.

\begin{figure}
\includegraphics[width=7cm,height=7cm,angle=-90]{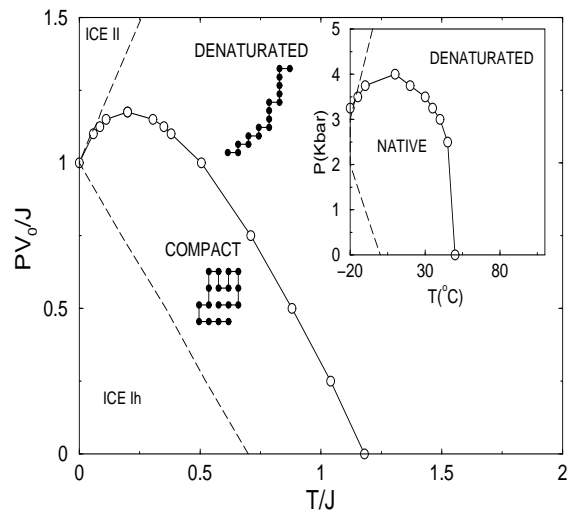}
\caption{P-T phase diagram for the protein derived from Fig.~\ref{fig1}.
  Dashed lines indicate the freezing lines for model water.  Water freezes in
  low density ice Ih for $PV_{0}/J<1$ and in dense ice II for $PV_{0}/J>1$.
  In the inset we present the experimental results obtained by Zhang et al.
  \cite{Zhang} for the bovine pancreatic ribonuclease A. Two typical
  configurations of the protein are shown, one in the compact state and the
  other in the denaturated state.  }
\label{fig2}
\end{figure}

We also reconstruct the phase diagram of the water-protein system in the
$P-T$ plane (Fig.~\ref{fig2}).  We consider the protein to be in the
collapsed state if $96\%$ of all possible contacts are formed, i.e., if
$\overline N_{c}/n_{max}>0.96$. For each pressure value, the freezing lines
of water shown in Fig.~\ref{fig2} are given by the temperature at which we
observe a maximum in the specific heat of the water bath.  We compare our
findings to experimental observations~\cite{Zhang} for bovine pancreatic
ribonuclease A studied by $^{1}$H NMR spectroscopy. We find remarkable
qualitative agreement between the experimental and numerical $P-T$ phase
diagrams. In both experimental and numerical $P-T$ phase diagrams, we observe
that cold denaturation occurs at high pressures and, as we lower the
temperature, close to the water--ice II freezing line and in the region where
water molecules are not capable of forming low density ice--like structures.
In addition to the study of ribonuclease A \cite{Zhang}, cold denaturation at
very high pressures in the kbar range has also been observed in
chymotrypsinogen~\cite{Hawley}, myoglobin \cite{Hawley}, staphylococcal nuclease
\cite{Panick99} and has been proposed as the principal mechanism for the
observed pressure-inactivation of bacteria, such as {\it Escherichia coli}
\cite{Ludwing96}.

\begin{figure}
\includegraphics[width=7cm,height=7cm,angle=-90]{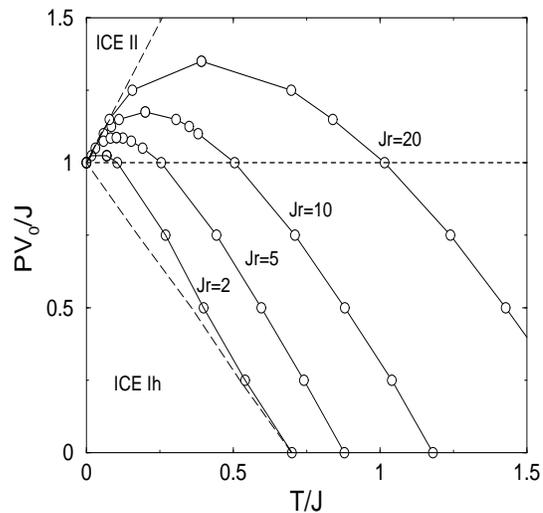}
\caption{P-T Phase diagram for proteins with $J_{r}=2,5,10,20$.
Dashed lines indicate the computed freezing lines for water.
Doted line indicates the $P=P_c=J/\Delta V$ line.
}
\label{fig3}
\end{figure}

Not all proteins behave equally as we decrease temperature at high pressure.
In particular, there are some proteins that do not exhibit cold denaturation.
We reproduce the variability of protein dynamics at high pressure and low
temperature by varying the hydrophobic parameter $J_{r}$ to lower values,
effectively impeding a stable compact state for pressures above the $P=P_c$
line.  In Fig.~\ref{fig3} we present the phase diagrams obtained for
different values of $J_{r}$, ranging from two to $20$. The shape of the phase
diagram changes as we increase the value of the repulsive interaction
$J_{r}$, allowing stabilization of the compact state and cold denaturation
above the $P=P_c$ line.

Within the framework of our model, we reproduce the experimentally-observed
{\it thermodynamics} of cold denaturation, but we cannot address the {\it
  kinetics} of this process. It could be also feasible to investigate in
future work the dynamics with a more sophisticated model, where we consider
not just the average number of hydrogen bonds with $\Delta V>0$, but the
actual numbers for each water molecule that is a neighbor to a residue.

Recent computer simulations studies \cite{Payne97} with all-atom models have
studied the effect of pressure and average density on the hydrophobic effect.
The authors performed simulations of two Lennard-Jones (LJ) particles in the
TIP4P water model \cite{Mahoney00}.  At constant temperature, they found that
the aggregation of the two particles is favored with a moderate increase of
pressure, or analogously, with a moderate increase of water density. However,
this effect is reversed for pressures in the kbar range, so that aggregation
becomes unstable with increasing water density.  These studies support our
results of a critical pressure above which an increase of pressure leads both
to an increase of water density (because of the reduction of the number of
hydrogen bonds with $\Delta V > 0$) and to destabilization of the model
protein. Following these results and based on previous studies
\cite{Chandler}, Shimizu et al. \cite{Shimizu02} have shown that three-body
interactions have a destabilization effect on the aggregation of three
LJ particles.  However, the inclusion of these interactions into a model of a
more complex protein did not lead to significant changes. Finally, all-atom
simulations recently addressed the pressure denaturation of proteins
\cite{Paci02}, but they have not provided conclusive evidence.

We conclude that the effect of pressure on water density is key for
understanding cold denaturation of proteins. The density anomaly of water
arises from the low density hydrogen bonded structures responsible for the
hydrophobic effect, driving the protein to a compact state
\cite{Privalov,DillII,Chandler,Frank}. At extreme pressures above $P_c$,
lowering the temperature implies an increasing free energy cost to form a
hydrogen bond with $\Delta V>0$, so the density anomaly disappears. In this
scenario, the hydrophobic effect decreases and cold denaturation occurs.  Our
model supports this mechanism. Also, a specific arrangement of amino acids in
the protein structure, determined by amino acid interactions, dictates
the dynamics of proteins at low temperature and high pressure, thus making
some proteins more stable than others at these $P-T$ conditions.

We thank S. V. Buldyrev, D. Chandler, J. Hermans, G. Hummer, O. Mishima and
F. Sciortino for helpful interactions, the NSF Chemistry Program Grant
CHE0096892, the Spanish Ministry of Education and the Petroleum Research Fund
for support.

% Create the reference section using BibTeX:
%\bibliography{water}

\begin{thebibliography}{99}
\bibitem{other} High pressure is not the only mechanism promoting cold
  denaturation. Alterations in the solvent conditions, such as variations in
  the pH \cite{Hawley, Cho82} or the addition of chemical denaturants
  \cite{Chen89}, may weaken the hydrophobic interactions, inducing cold
  denaturation of selected proteins at ambient pressures.
\end{thebibliography}

\end{document}